\documentclass[reprint,aps,prl,showpacs,floatfix,superscriptaddress,nofootinbib]{revtex4-1}

\usepackage{graphicx}
\usepackage{amsthm}   
 \usepackage[bottom]{footmisc}
\usepackage{amsmath} 

\usepackage{amssymb}  
\usepackage{mathrsfs} 
\usepackage{stmaryrd} 
\usepackage{txfonts} 

\usepackage{enumerate}
\usepackage{appendix}

\newcommand{\mathsym}[1]{{}}
\newcommand{\unicode}[1]{{}}

\newcommand{\E}{{\mathbb{E}}}

\newcommand{\F}{{\mathcal{F}}}
\newcommand{\V}{{\mathcal{V}}}
\newcommand{\G}{{\mathscr{G}}}
\newcommand{\WW}{\Omega}
\newcommand{\PP}{{\mathcal{P}}}

\usepackage{color}  
\RequirePackage[dvipsnames,usenames]{xcolor}

\newcommand{\be}{\begin{equation}}
\newcommand{\bel}[1]{\begin{equation}\label{#1}}
\newcommand{\qe}{\end{equation}}
\newcommand{\ee}{\end{equation}}
\newcommand{\eeq}{\end{equation}}
\newcommand{\ba}{\begin{eqnarray}}
\newcommand{\ea}{\end{eqnarray}}

\newcommand{\erf}[1]{Eq.~\eqref{#1}}

\date{\today}                      

\begin{document}

\title{Extending Landauer's Bound from Bit Erasure to Arbitrary Computation}

 \author{David H. Wolpert}
\altaffiliation{Massachusetts Insitutute of Technology}
 \altaffiliation{Arizona State University}
  \affiliation{Santa Fe Institute, 1399 Hyde Park Road, Santa Fe, NM 87501, USA\\
   \texttt{http://davidwolpert.weebly.com}}
%

%
%
%

\begin{abstract}
Recent analyses have calculated the minimal thermodynamic work
required to perform a computation $\pi$ when two conditions hold: the output of $\pi$ is independent of its
input (e.g., as in bit erasure); we use a  physical computer $\mathcal{C}$ to implement $\pi$ that is specially
tailored to the environment of $\mathcal{C}$, i.e., to the precise distribution over $\mathcal{C}$'s inputs, $\PP_0$.
First I extend these analyses to calculate the work required even if the
output of $\pi$ depends on its input, and even if $\mathcal{C}$ is not used with the distribution $\PP_0$ it was tailored for.
Next I  show that if $\mathcal{C}$ will be re-used, then
the minimal work to run it depends only on the logical computation $\pi$, 
independent of the physical details of $\mathcal{C}$.  
This establishes a formal identity between the thermodynamics of (re-usable) computers and
theoretical computer science. I use this identity to prove that the minimal work required to compute a bit
string $\sigma$ on a ``general purpose computer" rather than a special purpose one, i.e., on a universal Turing machine $U$, is
$k_BT\ln(2) \big[$Kolmogorov complexity($\sigma) \;+$
log (Bernoulli measure of the set of strings that compute $\sigma) \;+ $ log(halting
probability of $U)\big]$. I also prove that using $\mathcal{C}$ with a distribution over environments
results in an unavoidable increase in the work required to run the computer, even if it is tailored to that distribution over environments.
I end by using these results to relate the free energy flux incident
on an organism / robot / biosphere to the maximal amount of computation that the organism / robot / biosphere
can do per unit time.
\end{abstract}

\maketitle

There has been great interest for over a century in the relationship between thermodynamics and 
computation~\cite{szilard1964decrease,bril62,wiesner2012information,still2012thermodynamics,prokopenko2013thermodynamic,prokopenko2014transfer,zurek1989thermodynamic,zure89b,bennett1982thermodynamics,lloyd1989use,lloyd2000ultimate,del2011thermodynamic,fredkin1990informational,fredkin2002conservative,toffoli1990invertible,leff2014maxwell}.
A breakthrough was made with the argument of Landauer  
that at least $kT \ln[2]$ of work is required to run a 2-to-1 map like bit-erasure on any physical system~\cite{landauer1961irreversibility,landauer1996minimal,landauer1996physical,bennett1973logical,bennett1982thermodynamics,bennett1989time,bennett2003notes,maroney2009generalizing,plenio2001physics,shizume1995heat,sagawa2009minimal,dillenschneider2010comment,fredkin2002conservative},
a conclusion that is now being confirmed
experimentally~\cite{dunkel2014thermodynamics,roldan2014universal,berut2012experimental,koski2014experimental,jun2014high}. 
%
A related conclusion was that a 1-to-2 map
can act as a \emph{refrigerator} rather than a heater, \emph{removing} heat from the environment~\cite{bennett1989time,bennett2003notes,landauer1961irreversibility,bennett1982thermodynamics}.
For example, this occurs in adiabatic demagnetization of an Ising spin system~\cite{landauer1961irreversibility}.
%
%

This early work leaves many issues unresolved however. In particular,
say any output can be produced by our map, with varying
probabilities, from any input.  So the map is neither a pure heater nor a pure refrigerator.
What is the minimal required work in this case?
%

More recently, there has been dramatic progress in our understanding of
non-equilibrium statistical physics and its relation to information-processing~\cite{faist2012quantitative,touchette2004information,sagawa2012fluctuation,crooks1999entropy,crooks1998nonequilibrium,chejne2013simple, jarzynski1997nonequilibrium,esposito2011second,esposito2010three,parrondo2015thermodynamics,sagawa2009minimal,pollard2014second,seifert2012stochastic,dillenschneider2010comment,takara2010generalization,hasegawa2010generalization,prokopenko2015information,takara2010generalization}.
Much of this recent literature 
has analyzed the minimal work required
to drive a physical system's (fine-grained, microstate) dynamics during the interval from $t=0$ to $t=1$
in such a way that the dynamics of the macrostate is controlled by some desired Markov kernel $\pi$.
In particular, there has been detailed analysis of the minimal work needed 
when there are only two macrostates, $v =0$ and $v=1$, and we require that both get mapped to the bin $v=0$~\cite{esposito2011second,sagawa2014thermodynamic,parrondo2015thermodynamics}. By identifying the 
macrostates $v \in V$ as Information Bearing Degrees of Freedom (IBDF~\cite{bennett2003notes}) 
of an information-processing device like a digital computer, these
analyses can be seen as elaborations of the analyses of Landauer et al. on the thermodynamics of bit erasure.

Many of the work-minimizing systems considered in this recent literature proceed in two stages. First, they physically change an
initial, non-equilibrium distribution over microstates to the equilibrium distribution, $\rho^{eq}(w)$,
in a quenching process. 
All information concerning the initial microstate is lost from the distribution over $w$ by the end of this 
first stage. So in particular all information is lost about what the initial bin $v_0$ was. In addition, 
the Hamiltonian used in this quench is defined in terms of $\PP_0$, the initial distribution over computer inputs.
There is some unavoidable extra work if the computer is used with
an initial distribution that differs from $\PP_0$.

Next, in the second stage
$\rho^{eq}(w)$ is transformed to an ending (non-equilibrium) distribution over $w$, with an
associated distribution over the ending coarse-grained bin, $v_1$. However since 
all information about $v_0$ has been lost
by the beginning of the second stage, $v_0$ cannot have any effect on the distribution over $v_1$
produced in the second stage. Accordingly, changing the distribution over inputs to one of these systems has
no effect on the distribution over outputs. So although such a system can be used to implement a many-to-one map over the 
IBDF (i.e., the bins) in a digital computer, it cannot be used to implement any computational 
map whose output varies with its input.

In this paper I show how to implement any given conditional distribution $\pi$ with minimal work,
even if $\pi$ maps different initial macrostates $v_0$ to different final macrostates $v_1$. 
I do this by connecting the original, \textbf{processor} system with macrostates $v \in V$
to  a separate, initialized ``memory system" that records $v_0$, and then evolve the joint system in such a way that
the processor dynamics effectively samples $\pi(. \mid v_0)$. 
After this the memory is re-initialized (i.e., the stored copy of $v_0$ is erased), completing the cycle.

Like the systems considered in the
literature, those considered here are implicitly optimized for some ``prior" distribution over the inputs, $\G_0$.
Here I go beyond the analyses in the literature by allowing the {actual} distribution
over inputs, $\PP_0(v)$, to differ from our assumed distribution, $\G_0$. 
When $\G_0 = \PP_0$, the dynamics of the joint system is 
thermodynamically reversible. So the second law tells us that there is no alternative
system that implements $\pi$ with less work. However if $\G_0 \ne \PP_0$ (i.e., the computer
is used with a different user from the one they are optimized for) and $\pi$ is not just a permutation
over $v$, some of the work when the memory is reinitialized is unavoidably wasted.
I then analyze the situation where there is a distribution over $\PP_0$ (e.g., as occurs if the system is a computer that will be
used with multiple users, or if it is an organism that will experience different environments) 
and $\G_0$ is optimized for that distribution, deriving
how much extra work is needed due to uncertainty about who the user is.


I also show that \emph{if the physical system used to run the computation will be re-used}, then
the ``internal entropies", giving the entropy internal to each coarse-grained bin, do not contribute to the minimal work. 
In such a scenario
the specifics of the physical system implementing the computation --- which are reflected
in those internal entropies --- are irrelevant. The work depends only on
the computation $\pi$ implemented by that system. (In previous analyses the computer
was not re-used, so the internal entropies --- and therefore physical details of the computer --- were relevant.) This result
establishes a formal identity between the thermodynamics of (re-usable) computers
and theoretical computer science. 

As an illustration, I use this identity to analyze the thermodynamics of  
a ``general purpose computer" rather than a special purpose one, i.e., of a universal Turing machine $U$,
where the macrostates are labelled by bit strings. In particular I prove that the work required to compute a 
particular bit string $\sigma$ on $U$ is  $k_BT\ln(2)$ times the sum of the Kolmogorov complexity of $\sigma$, 
log of the Bernoulli measure of set of all strings that compute $\sigma$, and log of the
Halting probability for $U$. Intuitively, by considering \emph{all} input strings that result in
$\sigma$, the second term quantifies ``how many-to-one" $U$ is, something that is not captured
by the Kolmogorov complexity of $\sigma$.

I end by using these results  to relate the free energy flux incident
on an organism (robot, biosphere) to the maximal ``rate of computation" implemented by that
organism (resp., robot, biosphere).

I refer to the engineer who constructs the system as its ``designer", and 
refer to the person who chooses its initial state as its ``user". While the language of 
computation is used throughout this paper,
the analysis applies to any dynamic process $\pi$ over a coarse-grained partition of a fine-grained space, not
just those processes conventionally viewed as computers.
So for example, the analysis applies to the dynamics of biological organism reacting to its environment,
if we coarse-grain that dynamics; the organism is the ``computer", the dynamics is the ``computation",
the ``designer" of the organism is natural selection, and the ``user" initializing the organism is
the environment.

$ $

\noindent \emph{Problem setup} --- I write $|X|$ for the number of elements $x$ in any finite space $X$, and
write the Shannon entropy of a distribution $p$ over $X$ 
as $S_p(X) = S(p) =  -\sum_x p(x) \ln[p(x)]$, or even just $S(X)$ when $p$ is implicit. 
I use similar notation for conditional entropy, etc. 
I also write the cross-entropy between two distributions $p$ and $q$ both defined over some space $X$ as
$C(p(X) \mid\mid q(X)) \equiv -\sum_x p(x) \ln[q(x)] $
or sometimes just $C(p \mid \mid q)$ for short~\cite{coth91,mack03}. 


Let $W$ be the space of all possible microstates of a system
and ${\V}$ a partition of $W$, i.e., a coarse-graining of it into macrostates. For example, 
in a digital computer, $\V$ maps each microstate of the computer,
$w \in W$, into the bit pattern in the computer's memory.
I assume that the set of labels of the partition elements, $V$, contains ``0".
When convenient, I subscript a partition element with a time that the system state
lies in that element, e.g., writing, $v_0, v_1$, etc.

The Hamiltonian over $W$ at $t=0$ is $H^\varnothing_{sys}$, with associated equilibrium (Boltzmann) distribution
$\rho^{eq}$. For simplicity, I assume that $\forall v \in V$, at the two times $t=0$ and $t=1$, $Pr(w \mid v)$
is the same distribution, which I write as $q^v_{in}(w)$. (N.b., $q^v_{in}(w) = 0$ if $\V(w) \ne v$.) As in the analyses of computers 
in~\cite{bennett1989time,bennett2003notes,landauer1961irreversibility,bennett1982thermodynamics},
there is a ``user" of the system who intervenes in its dynamics at or before $t=0$, which 
results in the initial macrostate $v_0 \in V$ being set by sampling a \textbf{user distribution}
$\PP_0(v)$. As examples, $\PP_0$ could model randomness in how a single
user of a computer initializes the computer at $t=0$, 
or randomness in how an environment of an organism perturbs the organism at $t=0$.
I write the (potentially non-equilibrium) unconditioned distribution over $W$ at $t=0$ as
$
\PP_0(w) \equiv \sum_{v} \PP_0(v) q^{v}_{in}(w)
$.

The evolution of the microstates $w \in W$  during $t \in [0, 1)$ results in a conditional
distribution over macrostates, $\pi(v_1 \mid v_0)$.  
Since they are set by the designer of the system, I take $\pi$ 
and the distributions $q^{v}_{in}$ to be fixed and known to that designer.
However I allow the designer to be uncertain about what $\PP_0$ is.
As shorthand, I write 
$
\PP_1(v) \equiv \sum_{v} \PP_0(v) \pi(v \mid v)
$

I wish to focus on the component of the thermodynamic work that reflects
computation, ignoring the component that reflects physical labor. This is guaranteed if the expected 
value of the Hamiltonian at $t=0$ and $t=1$ is the same, regardless of $\PP_0$
and $\PP_1$, since that means that the change in the expected value of the Hamiltonian is zero. 
Accordingly I assume that at both $t=0$ and $t=1$, the expected value of the Hamiltonian if the 
system is in state $v$ then (i.e., $\sum_w  q^{v}_{in}(w) H^\varnothing_{sys}(w)$) is a
constant  independent of $v$. I write that constant as $h^\varnothing_{sys}$.
To simplify the analysis below, I also assume that $\E_{\rho^{eq}}[H^\varnothing_{sys}(w)] = h^\varnothing_{sys}$.

$ $

\noindent \emph{Overview of the system} ---  
The designer's goal is to modify the system considered in~\cite{esposito2011second,sagawa2014thermodynamic,parrondo2015thermodynamics} into one which no longer loses the
information of what the initial macrostate $v_0$ was as it evolves from $t=0$ to $t=1$.
%
%
This can be done by coupling the system with an
error-free \textbf{memory apparatus}, patterned after the measurement apparatus introduced
in~\cite{parrondo2015thermodynamics,sagawa2014thermodynamic,sagawa2009minimal}. As in those studies,
the ``measurement" is a process that copies the macrostate
to an initialized, external, symmetric memory with the value of $v_0$, and does so without changing $v_0$ (or even 
the initial microstate of the processor, $w_0$). Having set the value of such a memory,
we can use its value later on, to govern the dynamics of $w$ after the time when the
the system distribution has relaxed to $\rho^{eq}(w)$, to ensure that $v$ evolves according
to $\pi$ --- even if under $\pi$ the ending macrostate of the system depends on its initial state.
Finally, to complete a cycle, the memory apparatus must be reinitialized.


I assume that the memory and system are both always in contact with a heat bath at temperature $T$. 
To be able to store a copy of any $v \in V$, the memory must have the same set of possible macrostates, $V$.
I write the separate memory macrostates as $m \in V$, with associated microstates $u \in U$. (\emph{A priori}, 
$U$ need not have any relation to $W$.) 
For simplicity I assume that the conditional distribution of $u$ given any $m$ is the same distribution $Q^m_t(u)$ at both $t=0$
and $t=1$, and that there is a uniform equilibrium Hamiltonian $H^\varnothing_{mem}(u)$.
In addition, I make the inductive hypothesis that the starting value of the memory is $m = 0$, with probability 1.
The system dynamics comprises the following four steps (see Fig.~\ref{fig:measure_evolve}):
 \noindent
 \begin{figure}[tbp]
        \includegraphics[width=\textwidth]{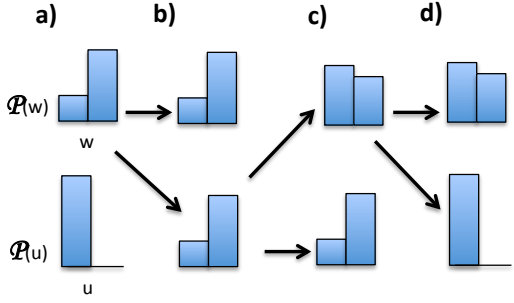}
	\vglue-75mm
	\hglue-20mm
        \noindent \caption{Example of dynamics of the marginal distributions of a system with a binary coarse-graining, where
        the bins have the same size and both $q^v_{in}(w)$ and $Q^m(u)$ are uniform for all $v$, $m$. The top row shows
        the dynamics of the processor, and the bottom row shows the dynamics of the memory.
        Fig.~(a) shows the $t=0$ state, with the right bin of the processor more probable than the left bin, and the memory is in
        its initialized bin. Fig.~(d) shows the $t=1$ state, where the relative probabilities of the processor bins have changed
        according to $\pi$, and the memory has been returned to its initialized bin. 
                  }
 \label{fig:measure_evolve}
 \end{figure}
 
\noindent \emph{I --- }  First the memory apparatus copies the initial value $v = v_0$
into the memory, i.e., sets $m = v_0$. This step
is done without any change to $w$, and so $\PP_0(w)$ is unchanged. Since the copy is error-free and the 
memory is symmetric, this step does not require thermodynamic work~\cite{parrondo2015thermodynamics}. 

\noindent \emph{II --- } Next a Quench-then-Relax procedure (QR) like the one described in~\cite{parrondo2015thermodynamics,esposito2011second} is run on the distribution over $w$,
$q^{v_0}_{in}(w)$.
In such a QR, first we replace $H^\varnothing_{sys}$ with a \textbf{quenching Hamiltonian} 
chosen such that $q^{v_0}_{in}$ is an equilibrium distribution for a Hamiltonian specified by
the memory system macrostate:
\ba
H^{m}_{in}(w) &\equiv& -kT \ln[q^{v_0}_{in}(w)]
\label{eq:quenching_ham_1}
\ea
(While $w$ is unchanged in this adiabatic quench, and therefore so is the distribution over $W$,
in general work is required if $H^{m}_{in} \ne H^\varnothing_{sys}$.)
Next we isothermally and quasi-statically relax $H^{m}_{in}$ back to $H^\varnothing_{sys}$, thereby changing 
$q^{v_0}_{in}(w)$ to $\rho^{eq}(w)$. (See also~\cite{sagawa2014thermodynamic}.)

\noindent \emph{III ---} Next we use the fact that $m = v_0$ to run a QR over $W$ in reverse, with
the quenching Hamiltonian 
\ba
H^{m}_{out}(w) &\equiv& -kT \ln[q^{v_0}_{out}(w)] 
\ea
where $q^{v_0}_{out}(w) \equiv \sum_{v_{1}} q^{v_{1}}_{in}(w) \pi(v_{1} \mid v_0)$.
This reverse QR begins by isothermally
and quasi-statically sending $H^\varnothing_{sys}$ to $H^{m}_{out}$. After that $H^{m}_{out}$ is replaced by 
$H^\varnothing_{sys}$ in a ``reverse quench", with no change to $w$. As in step (II), there is no change to $m$ in step (III).

\noindent \emph{IV --- } Finally, as described in detail below, we reset $m$ to 0. This ensures we can rerun 
the system, and also guarantees the inductive hypothesis.

Since the system samples $\pi(v_1 \mid v_0)$ in step (III), 
these four steps implement the map $\pi$ even if $\pi$'s output
depends on its input, and no matter what $\PP_0$ is. (The whole
reason for storing $v_0$ in $m$ was to allow this step III.) 

%
%
%
Moreover, the expected work expended in the first three steps is given (with abuse of notation) by the conditional entropy,
\ba
   -kT \bigg( S_{\PP}(V_{1} \mid V_0) + \sum_v S(q^{v}_{in}) \bigg[\PP_1(v) - \PP_0(v) \bigg] \bigg)
\label{eq:change_v}
\ea
(See Supplementary Material (SM) 
for proof.)

$ $

\noindent \emph{Resetting the memory} --- We implement step (IV)
by first running a QR on the distribution over $U$ (not $W$), and then running a reverse
QR, one that ends with $m = 0$ no matter what the initial value of $m$ was.
%

In detail, suppose that the designer of the system guesses that the distribution over the initial values of the macrostates
is $\G_0(v_0)$ --- which in general need not equal $\PP_0$.
This distribution would be the prior probability over the values of $m$ if $G_0$ equalled  $\PP_0$, since $m$ is a copy of $v_0$.
The associated likelihood of $v_{1}$ given
$m$ is $\G(v_1 \mid m) = \pi(v_{1} \mid v_0 = m)$. So the posterior probability of $m$ given $v_{1}$, $\G(m \mid v_{1})$, is
proportional to  $\G_0(m) \pi(v_{1} \mid m)$. This gives the (guessed) posterior probability over memory microstates,
which we can write as
\ba
\G(u \mid v_{1}) &=&  \sum_{m} \G(m \mid v_{1}) Q^{m}(u)
\ea
with some abuse of notation.
In contrast, the actual posterior distribution $\PP(m \mid v_{1})$ is given by the actual prior $\PP_0$, and gives
a posterior distribution
\ba
\PP(u \mid v_{1}) &=& \sum_{m} \PP(m \mid v_{1}) Q^{m}(u)
\ea

The premise of this paper is that to reset the memory the computer first runs a QR
using the quenching Hamiltonian 
\ba
H^{v_{1}}_{mem}(u) \equiv -kT \ln \G(u \mid v_{1})  
\ea
to drive the distribution over $u$ to $\rho^{eq}_{mem}(u)$,
since this would relax the memory using minimal work if the guessed prior $\G_0$ equaled the actual one, $\PP_0$.
(Intuitively, $v_{1}$ is a ``noisy measurement" of $m$
that is used to set this quenching Hamiltonian, and we are running
the same process as in step II, just with the roles of the memory and
processor reversed.)
Next we run a reverse QR,
taking $\rho^{eq}_{mem}(u)$, the uniform 
distribution over all $U$, to the distribution that is uniform over $m = 0$, zero elsewhere. 
This completes the resetting of the memory macrostate.

Averaging the work required in this resetting of the memory,
and adding it to the expression in~\erf{eq:change_v}, gives
the minimal expected work for running $\pi$:
\ba
\!\!\!\! \!\!\!\! \!\!\!\! \WW_{\G_0, \PP_0} &\equiv& -kT \bigg( \sum_{v_{0}, v_{1}} \PP(v_0, v_{1}) \ln \big[ \G(v_0 \mid v_{1})  \big] \nonumber \\
   &&+ \; S_{\PP}(V_{1} \mid V_0)   \;+\; \sum_v S(q^v_{in}) \bigg[\PP_1(v) - \PP_0(v) \bigg] \bigg)
\label{eq:final_answer_1}
\ea
(See SM for proof.)
Since $\G(v_1 \mid v_0) = \pi(v_1 \mid v_0) = \PP(v_1 \mid v_0)$,
we can use Bayes' theorem to rewrite~\erf{eq:final_answer_1} as
\ba
&& \!\!\!\!\!\!\!\!\!\! kT  \bigg( C [\PP(V_0) \mid\mid \G(V_0))] - C [\PP(V_1) \mid\mid \G(V_1))]  \nonumber \\
&&  \qquad \qquad \qquad  \qquad + \sum_v S(q^v_{in}) \bigg[\PP_0(v) - \PP_1(v) \bigg]  \bigg)
\label{eq:final_answer_2}  
\ea
(assuming all distributions over $V$ have the same support, so that 
we don't divide by zero when using Bayes' theorem).

So if $\G_0 = \PP_0$, or alternatively $\pi$ is an invertible function over $V$,
$ \WW_{\G_0, \PP_0} = kT [
S_{\PP_0}(W) - S_{\PP_1}(W)]$. This quantity is sometimes called ``generalized Landauer cost". Note that for
a fixed $P(w_0, w_1)$, it is independent of the partition.


$ $

\noindent \emph{Multiple cycles of a computer} --- 
Sometimes we will want to use an (IID) \textbf{calculator} computer, in
which we IID sample $\PP_0$ at the end of each iteration, over-writing $v_1$, before running $\pi$ again. 
In such calculators, after step (IV) above, the value $v_1$ is copied
to an external system via an additional memory apparatus (e.g., in order to drive some physical actuator). 
Then a different external system (e.g., a sensor) forms a sample $v'_0 \sim \PP_0$, and 
$v_1$ gets replaced by $v'_0$. Only after these two new steps have we completed a full cycle. At this point we can run 
another cycle, to apply $\pi$ again --- but starting from $v'_0$ rather than $v_1$. 

In the SM it is shown that for an ``extended" calculator computer, where $\pi$ is iterated $N$ times and only then is $v$
copied to an external system and $v'_0$ copied in, the total work expended is at least
\ba
&& \!\!\!\!\!\!\!\!\!\! kT  \bigg( C [\PP(V_0) \mid\mid \G(V_0)] - C [\PP(V_N) \mid\mid \G(V_N)]  \bigg) \; 
%
\label{eq:calculator_work}	
\ea


Note that the expected work of a calculator has no dependence on the values $S(q^v_{in})$; 
in calculators the work depends only the logical map that
$\pi$ implements over $V$, independent of the physical system that implements that map.
So there is a formal identity between the thermodynamics of (calculator) computers and computer 
science.

As an example, fix a prefix-free universal Turing machine $U$~\cite{hopcroft2000jd,livi08,grunwald2004shannon}.
Identify the macrovariable $v \in V$ of the physical system implementing $U$ with the 
instantaneous description (ID) of $U$, so that $\pi$ gives the dynamics over those IDs,
 i.e., it is the dynamical law implementing the Turing machine.  I will say that an instantaneous description 
 (ID) of $U$ is a  \textbf{starting} ID if it specifies that the machine $U$ is in its initial state with 
an input string for which $U$ halts. Also define
$I^\sigma$ as the set of all starting IDs that halt with output $\sigma$,
and $K_U(\sigma)$ as the Kolmogorov complexity of $\sigma$. Finally,
for any starting ID $\alpha$, define $\ell(\alpha)$ as the length of the input string of $\alpha$.

Perhaps the most natural prior probability distribution 
over IDs is the (normalized) ``universal prior probability", i.e., $\G_0(v) = 2^{-\ell(v)} / \lambda$ if $v$ is
a starting ID and $\G_0(v) = 0$ otherwise, where $\lambda$ is $U$'s halting probability.
It is shown in the SM that under the simpler of two  
natural definitions of how to use a TM $U$ as a ``calculator computer", the minimal work (over all possible $\PP_0$(
needed to compute $\sigma$ 
is 
\ba
\!\!\!\!\!\! 
 kT \ln(2) \bigg(K_U(\sigma) + \log[\G_0(I^\sigma)] + \log \lambda  \bigg)
\label{eq:kolmogorov}
\ea

So 
the greater the gap between the log-probability that a randomly chosen program computes $\sigma$ and the log-probability of the 
most likely such program, the greater the work to compute
$\sigma$. Intuitively, running $U$ on $I^\sigma$ executes a many-to-one map in the Landauer 
sense, taking many starting IDs to the same ending ID. 
The gap between $\log[\G_0(I^\sigma)]$ and min$_{v_0 \in I^\sigma}\log[\G_0(v_0)] = K_U(\sigma) + \log \lambda$
quantifying ``how many-to-one" that map is. (Similar results hold for other choices of space of logical
variables $V$, machine $\pi$ and / or prior $\G_0$.) 

As an aside, by Levin's coding theorem~\cite{livi08},
$K_U(\sigma) + \log[\G_0(I^\sigma)]$ is bounded by a constant that depends only on $U$, and is independent
of $\sigma$. So for any $U$, there is a $\sigma$-independent upper bound on the minimal amount of
work needed for $U$ to compute $\sigma$.

$ $

\noindent \emph{Multiple users} --- Often rather than a single user of a calculator computer there will be a 
distribution over users, $Pr(\PP_0)$. To analyze this situation, 
use ~\erf{eq:final_answer_1} to write
\ba
\langle \Omega_{\G_0, \PP_0} \rangle &=& \Omega_{\G_0, \langle \PP_0 \rangle}
\ea
(where $\langle . \rangle$ indicates an average according to $Pr(.)$).
Applying this equality to~\erf{eq:calculator_work}, and using the facts that
Kullbach Leibler (KL) divergence is non-increasing in $t$ and is minimized (at zero) when its arguments are equal~\cite{coth91},
we see that the $\G_0$ that minimizes expected work is $\langle \PP_0 \rangle$.
The associated expected work is $S_{{\langle{\PP}\rangle}}(\V_0) - S_{{\langle {\PP} \rangle}}(\V_{1})$.

The expected work would instead be $\langle S_{{{\PP}}}(\V_0) - S_{{{\PP}}}(\V_{1})\rangle$ 
if we could somehow re-optimize $\G_0$ for each $\PP_0$. So 
the difference between those two values of expected work can be viewed as the minimal penalty we must pay 
due to uncertainty about who the user is. This penalty can be re-expressed as the drop from $t=0$ to $t=1$ in
the \textbf{entropic variance}, 
\ba
 \langle \PP \ln[\PP] \rangle - \langle  \PP \rangle \ln[\langle \PP \rangle ] 
\ea
i.e., it is the growth from $t=0$ to $t=1$ in certainty about $\PP$.

Entropic variance is non-negative and non-increasing.{\footnote{Resp., since entropy is a concave function of distributions,
and since entropic variance is the average (over $\PP_0$'s) of the KL divergence between 
$\PP$ and $\langle \PP \rangle$.}} So 
the work penalty that arises
due to growth in certainty about $\PP$ is always non-negative. This is true even if the minimal
work required to implement the  underlying computation is negative.

$ $

\noindent \emph{Implications for biology.} --- Any work expended on the processor must 
first be acquired as free energy from the processor's environment. However in
many situations there is a limit on the flux of free energy through a processor's immediate environment.
Combined with the analysis above, such limits provide upper bounds on the ``rate of (potentially
noisy) computation" that can be achieved by a biological organism in that environment.
In particular, since the minimal work required to do a computation increases if $\G_0 \ne \PP_0$, using the same
biological organism in a new environment, differing from the one it is tailored for, will in general result in
extra required work.

As an example, these results bound the rate of computation of a human brain. 
Given the fitness cost of such computation (the brain uses $\sim$ 20\% of
the calories used by the human body), this bound contributes to the natural selective pressures on humans,
in the limit that operational inefficiencies of the brain have already been minimized.
In other words, these bounds suggest that natural selection imposes a tradeoff between the fitness quality of a brain's decisions,
and how much computation is required to make those decisions. In this regard, it is 
interesting to note that the brain is famously noisy --- and as discussed above, noise in
computation reduces the total thermodynamic work required. 

As a second example, the rate of solar free energy incident upon the earth 
provides an upper bound on the rate of computation that can be achieved by the biosphere.
(This bound holds for any choice for the partition of the biosphere's fine-grained space
into macrostates such that the dynamics over those macrostates executes $\pi$.)
In particular it provides
an upper bound on the rate of computation that can be achieved by human civilization, if we remain
on the surface of the earth, and only use sunlight to power our computation. 



\acknowledgements{\emph{Acknowledgements} ---
I would like to thank Josh Grochow, Cris Moore, Daniel Polani, Simon DeDeo, Damian Sowinski, Eric Libby, 
and especially Sankaran Ramakrishnan for many stimulating discussions,  
and the Santa Fe Institute for helping to support this research. This paper was made possible through the support of Grant No. TWCF0079/AB47 from the Templeton World Charity Foundation and Grant No.
FQXi-RHl3-1349 from the FQXi foundation. 
The opinions expressed in this paper are those of the author and do not necessarily 
reflect the view of Templeton World Charity Foundation.}

\bibliographystyle{amsplain}


\providecommand{\bysame}{\leavevmode\hbox to3em{\hrulefill}\thinspace}
\providecommand{\MR}{\relax\ifhmode\unskip\space\fi MR }
\providecommand{\MRhref}[2]{%
  \href{http://www.ams.org/mathscinet-getitem?mr=#1}{#2}
}
\providecommand{\href}[2]{#2}

\newpage

\section{Derivation of Eq. 3 of main text}


In this section I evaluate the expected work required to implement the first three steps of the system for initial distribution
$\PP_0$. To do this,
it will be convenient to calculate the expected work to perform those steps conditioned on a particular $v_0$, and then 
average over all $v_0$ according to $\PP_0(v_0)$.

As in~\cite{parrondo2015thermodynamics}, I assume that after step (I) the interaction
Hamiltonian between $W$ and $U$ is negligible. Also as in that work, I assume that
the quench step at the beginning of step II is an instantaneous change to the 
energy of every $w$, $\Delta E(w)$. This process does not actually change $w$ (such changes
are associated with transfer of heat). Since the
quenching Hamiltonian depends on the value of $m$ (which due to step (I) depends on $v_0$), the value of
$\Delta E(w)$ for each $w$ also depends on $m$. That change in the
energy of $w$ is identified as the work done on the system in the quench step when it 
starts (and stays) in that state $w$. 

Now due to the fact that step (I) did not change $w$, at the beginning of the quench step
the posterior probability of $w$ given a current value $m$ is $q^{m}_{in}(w)$. Therefore
the expected work done in this quench step conditioned on a particular value $m$ is
$ \sum_{w} q^{m}_{in}(w) [H^{m}_{in}(w) - H^\varnothing_{sys}(w)]$. As shorthand,
define $S^0_{sys}$ as the Shannon entropy over $W$ for the Boltzmann distribution
with temperature $T$ and Hamiltonian $H^\varnothing_{sys}$.
Then conditioned on a value $v_0$ at the beginning of step (I), the work to perform the entire QR in step (II) 
is
\ba
 \sum_{w} q^{m}_{in}(w) [H^{m}_{in}(w) - H^\varnothing_{sys}(w)] + \F(H^\varnothing_{sys}) - \F(H^{m}_{in})
\label{eq:work_step_2}
\ea
where $m = v_0$ and
\ba
\F(H^\varnothing_{sys}) &=& h^\varnothing_{sys} - kT S^0_{sys}
\ea
is the equilibrium free energy of $H^\varnothing_{sys}$ at temperature $T$. 

By definition of 
$H^{m}_{in}$, $\F(H^{m}_{in}) = 0$. So the
expression in~\erf{eq:work_step_2} just equals $kT\bigg[S(q^{m}_{in}) - S^0_{sys}  \bigg]$.
Note that this amount of work is negative, since work is extracted
by sending $q^{m}_{in}$ to the equilibrium distribution for $H^\varnothing_{sys}$.

Similarly, to implement step (III) requires work of at least $kT\bigg[S^0_{sys} - S(q^{m}_{out})\bigg]$.{\footnote{In steps II and III
the usual convention was followed by quasi-statically sending $H^{m}_{in}$ to $H^\varnothing_{sys}$
and then sending $H^\varnothing_{sys}$ to $H^{v_0}_{out}$. 
The same total work would arise if we instead quasi-statically send $H^{m}_{in}$ to $H^{v_0}_{out}$ directly.}}
Now for any distribution $Pr(w)$, with some abuse of notation, we can write $S_{Pr}(v \mid w) = 0$, since $w$ sets $v$ uniquely.
Therefore 
\ba
S_{Pr}(w) &=& S_{Pr}(v \mid w) + S_{Pr}(w) \nonumber \\
  &=& S_{Pr}(v, w) \nonumber \\
   &=&  S_{Pr}(v) + S_{Pr}(w \mid v) 
\label{eq:trick}   
\ea
So if we write the Shannon entropy of the distribution over values $v_1$ conditioned on a particular
value of $v_0$ as
\ba
S_\pi(V_1 \mid v_0) &\equiv& -\sum_{v_1} \pi(v_1 \mid v_0) \ln[\pi(v_1 \mid v_0)]
\ea
then we can write 
\ba
S(q^{v_{0}}_{out}) &=& S_\pi(V_1 \mid v_0) - \sum_{w_1, v_1} \pi(v_1 \mid v_0) q^{v_1}_{in}(w_1) \ln(q^{v_1}_{in}(w_1)) \nonumber \\
  &=&  S_\pi(V_1 \mid v_0) + \sum_{v_1} \pi(v_1 \mid v_0)S(q^{v_1}_{in}) 
\ea

Accordingly, the total amount of work in the first three steps, conditioned on a value $v_0$, is
\ba
&& \!\!\!\!\! \!\!\!\!\! kT \bigg[S(q^{v_0}_{in}) - S(q^{v_0}_{out}) \bigg] \nonumber \\
   && = \; kT \bigg[S(q^{v_0}_{in}) - S_\pi(V_1 \mid v_0) - \sum_{v_1} \pi(v_1 \mid v_0) S(q^{v_1}_{in}) \bigg]
\ea
Combining and averaging under $\PP_0(v_0)$, the expected work required to complete the first three steps is
\ba
-kT \bigg[S_\pi(V_1 \mid \V_0) + \sum_{v} S(q^{v}_{in}) \bigg(\PP_1(v) - \PP_0(v) \bigg) \bigg]
\ea
(The analogous expression in much of the literature has $S_\pi(V_1)$
instead of $S_\pi(V_1 \mid V_0)$; the difference is due to the requirement that $\pi$ govern
the coarse-grained dynamics even if its output depends on its input, a requirement that means
that we must measure the value $v_0$.)


\newpage

\section{Derivation of Eq. 7 of the main text}

The QR in resetting the memory is run at $t=1$, using $H^{v_{1}}_{mem}(u)$.
It does not change $w_{1}$, just as measurement of $v_0$ did not change $w_0$.
Accordingly, the minimal amount of work in this QR  is
\ba
&& \!\!\! \sum_{u} \PP(u \mid v_{1})  [H^{v_{1}}_{mem}(u) - H^\varnothing_{mem}(u)] + \F(H^\varnothing_{mem})
 \nonumber \\
   && \; \; = \; kT \bigg(-\sum_{u} \PP(u \mid v_{1}) \ln \bigg[ \G(u \mid v_{1})  \bigg]
     - \ln|V| \bigg)
\label{eq:QR_mem_reset}     
\ea

This is true whether or not $\G_0 = \PP_0$. Note 
though that due to the fact that $H^{v_1}_{mem}$ is defined in terms of $\G_0(u \mid v_{1})$ not $\PP_0(u \mid v_{1})$,
if both $\G_0 \ne \PP_0$ and $\pi$ is not an invertible deterministic map, then the actual posterior $\PP(u \mid v_{1})$ is not the
equilibrium distribution for $H^{v_{1}}_{mem}$. This means that immediately after the quenching process,
as the Hamiltonian over $U$ begins to quasi-statically relax, the distribution over $U$ will first settle,
in a thermodynamically irreversible process, to the equilibrium distribution for $H^{v_{1}}_{mem}$. No work is
involved in that irreversible process. However if we had instead chosen $\G_0 = \PP_0$, the expression in~\erf{eq:QR_mem_reset} 
would have been less, i.e.,
less work would have been required, since no such irreversible process would have occurred.

To complete resetting the memory we now run
a reverse QR that takes $u$ from the uniform 
distribution over all $U$ to the distribution $Q^0(u)$, whose support is restricted
to $u$'s such that $m = 0$.
This means that for the given value of $v_1$, the total work required to reset $m$ to $0$,
including the contribution evaluated in~\erf{eq:QR_mem_reset}, is
\ba
&& \!\!\!\! -kT \bigg(\sum_{u} \PP(u \mid v_{1}) \ln \bigg[ \G_0(u \mid v_{1})  \bigg] \;+\; S(Q^0)\bigg)
\ea
Multiply and divide the argument of the logarithm in the summand by $\PP(u \mid v_{1})$.
Next use the same kind of decomposition
as in~\erf{eq:trick}, and then use the chain-rule for KL divergence. This transforms our
expression into
\ba
-kT \bigg( \sum_{v_{0}} \PP(v_0 \mid v_{1}) \ln \bigg[ \G_0(v_0 \mid v_{1})  \bigg]  - 
						\sum_{v_0}\PP(v_0 \mid v_1) S(Q^{v_0}) + S(Q^0) \bigg) \nonumber \\
\ea

Averaging this according to $\PP_{1}(v_{1})$ gives
\ba
-kT \bigg( \sum_{v_{0}, v_1} \PP(v_0, v_{1}) \ln \bigg[ \G_0(v_0 \mid v_{1})  \bigg]  - 
						\sum_{v_0}\PP(v_0) S(Q^{v_0}) + S(Q^0) \bigg) \nonumber \\
\label{eq:expected_reset_work}
\ea
Note though that we assumed that the states of the memory are symmetric. (This is why there is
no expected work in step (I).) So $S(Q^v)$ is independent of $v$, and~\erf{eq:expected_reset_work} reduces to
\ba
-kT  \sum_{v_{0}, v_1} \PP(v_0, v_{1}) \ln \bigg[ \G_0(v_0 \mid v_{1})  \bigg] 
\label{eq:expected_reset_work_1}
\ea
Adding~\erf{eq:expected_reset_work_1} to~\erf{eq:change_v} of the main text  
gives~\erf{eq:final_answer_1} of the main text, as claimed.

\newpage

\section{Derivation of Eq. 9 of main text}

Since no work is required in the new step where we measure $v_1$,
the total work in an iteration is given by adding~\erf{eq:final_answer_2}
to the additional average work required to map $v = v_1$ to $v = v'_0$. 
Since both the values $v_1$ and $v'_0$ exist outside of $W$,
they can be used to specify the two quenching Hamiltonians that implement this map. So
the additional average work is  
$
kT \sum_{v, v'} \bigg[S(q^{v}_{in})  \PP_1(v) - S(q^{v'}_{in})  \PP_0(v') \bigg].
$
Generalizing this reasoning gives
\ba
&& \!\!\!\!\!\!\!\!\!\! kT  \bigg( C [\PP(V_0) \mid\mid \G(V_0)] - C [\PP(V_N) \mid\mid \G(V_N)]  \bigg) \; 
%
\ea
as claimed. 

Note that this 
result requires the computer to contain an integer-valued clock, whose state $t$ increases
by $1$ at each iteration. This clock is needed so that the appropriate posterior $\G(m_t \mid v_{t+1})$ can be used to set
$H^{v_{t+1}}_{mem}(u_t)$ at iteration $t$. Note that such a clock can be implemented
without any work, since its dynamics is logically reversible. Given such a clock, 
the cross-entropies and internal entropies over iterations $t \in 2, \ldots, N-1$ cancel out.

\newpage

\section{Derivation of Eq. 10 of main text}

We are ultimately interested in the map from $U$'s input tape
to its output tape. In addition, $U$ is a prefix machine, i.e., its read tape
head cannot move to the left.
This motivates defining the IDs of $U$  as all tuples of \{machine state, contents of output tape, contents of work tape(s),
and contents of input tape \emph{at or to the right of the input tape read head up to the end of the
prefix codeword on the input tape}\}.

In addition, I require that $U$ can only halt if it has reinitialized its work tape(s). So all
IDs with $U$ in its halt state and output tape containing $\sigma$ have the same
(blank) work tape(s). This means that when $U$ halts there is no ``relic" 
recorded in the work tape(s) of what the original
contents of the input tape was. In addition, by the precise definition above of
IDs, all IDs with $U$ in its halt state and output tape containing $\sigma$ have no
information concerning the contents of the input tape. So
there is a unique 
ID with $U$ in its halt state and output tape containing $\sigma$; it does not matter
what input string to $U$ was used to compute $\sigma$. 

To simplify notation, let $f$ be the transition function of $U$, i.e., write
$\pi(v' \mid v) = \delta_{v', f(v)}$.  
Iterating $f$ from a starting ID $v_0$ eventually
results in an ID $v'$ that specifies that $U$ has halted. That $v'$ is
a fixed point of $U$. Write $\phi(v_0)$ for that fixed point arising
from the ID $v_0$ (i.e., $\phi$ is the partial function computed by $U$).
Also write $N(v_0)$ for the iteration at which $U$ halts (with output value $\phi(v_0)$). 
Finally, define $I^\sigma$
as the set of starting IDs that compute $\sigma$. Since
we are interested in user distributions $\PP_0$ that are guaranteed to compute $\sigma$, from now
on I restrict attention to $\PP_0$ whose support lies within $I^\sigma$. 

There are several ways to define ``the expected work for $U$ to compute $\sigma$"
using a calculator computer.  
In two of the most natural approaches, for any specific $v_0 \in I^\sigma$, the computer
is run some number of iterations, after which $v$ gets copied to the actuator and then reset,
and the total amount of work is tallied.
Where these approaches differ is in their rules for ``when $v$ gets copied to the actuator and reset".

In one approach we start with some specified $v_0 \in I^\sigma$ and then
\begin{enumerate}
\item Run the computer until it halts (with output $\sigma$) at timestep $N(v_0)$;
\item Copy that ending $v$ (which is just $\sigma$) to the actuator;
\item Set $v$ to its next value, $v'_0$, copied over from the sensor; 
\item Cease to exist. 
\end{enumerate}
In this approach, an iteration of the calculator is identified as the sequence
of iterations of $f$ that takes $v_0$ to a halt state. So different $v_0$ will
be identified with different numbers of iterations of $f$.

A second approach is the same as this first approach, except that we replace step (1) in this list
with iterating $f$ starting from the specified $v_0 \in I^\sigma$ a total of $\tau$ times.
We then consider the limit as $\tau \rightarrow \infty$. Since
$v_0$ is a starting state, we are guaranteed that under this limit, when we reach step (4) the 
computer has halted, and the value $\sigma$ has been
copied to the actuator. Moreover, as shown below, the minimal amount of work expended converges under this limit.

To evaluate the expected work in the first approach, 
combine the fact that $\pi$ is a deterministic function,~\erf{eq:calculator_work}, and the restriction
on the support of $\PP_0$ to write
\ba
&&-\sum_{v \in I^\sigma} \PP_0(v) \ln[\G_0(v)] + \sum_{v \in I^\sigma} \PP_{N(v)}(\phi(v)) \ln \bigg[\G_{N(v)} (\phi(v))\bigg]  \nonumber \\
&&  \qquad =  \;-\sum_{v \in I^\sigma} \PP_0(v) \ln[\G_0(v)] + \sum_{v \in I^\sigma} \PP_0(v) \ln \bigg[\G_{N(v)} (\phi(v))\bigg]  \nonumber \\
  && \qquad =  \;-\sum_{v \in I^\sigma} \PP_0(v) \ln[\G_0(v)] +  \sum_{v \in I^\sigma} \PP_0(v) 
  		\ln \bigg[\sum_{v' : f^{N(v)}(v') = \phi(v)} \G_0(v') \bigg]      			\nonumber \\
   && \qquad =  \; \sum_{v \in I^\sigma} \PP_0(v)  \ln \bigg[ \frac{\sum_{v' : f^{N(v)}(v') = \phi(v)} \G_0(v')}{\G_0(v)} \bigg]      			
\ea
So the optimal $\PP_0$ is a delta function about the $v \in I^\sigma$ that minimizes
\ba
\frac{\sum_{v' : f^{N(v)}(v') = \phi(v)} \G_0(v')}{\G_0(v)} &=& \frac{\sum_{v' : f^{N(v)}(v') = \phi(v)} 2^{-\ell(v')}}{2^{-\ell(v)}}
\ea
and the associated minimal amount of work is
\ba
&& \!\!\!\! kT \ln(2) {\mbox{min}}_{v \in I^\sigma} \bigg[ \ell(v) + \log \bigg( \G_0 \big(\{v' : f^{N(v)}(v') = \phi(v)\} \big) \bigg)
		+ \log \lambda \bigg]  \nonumber \\
   && \;\; = \; kT\ln(2) {\mbox{min}}_{v \in I^\sigma} \bigg[ \ell(v) + \log \bigg( \sum_{v' : f^{N(v)}(v') = \phi(v)} 2^{-\ell(v')} \bigg)
   			 \bigg]    
\label{eq:first_approach}   
\ea
where $\lambda$ is the normalization constant for Chaitin's omega, i.e., the halting probability
for $U$. 

Intuitively, in this first approach, the amount of work for computing $\sigma$ from some
$v \in I^\sigma$ is given by the difference of two terms. The first is the length of $v$, i.e., how unlikely $v$ is under $\G_0$.
Or to put it another way, it is ``how much information" there is in the initial ID of $U$. The second term is `how much information"
how much information there is concerning the initial ID of $U$ by the time the computation ends. The bigger the drop in the amount
of information concerning the initial ID, the more work is required to compute $\sigma$ from $v$. 

In contrast, in the second approach, the analogous analysis shows that the expected work is
\ba
&&-\sum_{v \in I^\sigma} \PP_0(v) \ln[\G_0(v)] + \lim_{\tau \rightarrow \infty} \bigg\{
    \sum_{v \in I^\sigma} \PP_{\tau}(\phi(v)) \ln \bigg[\G_{\tau} (\phi(v))\bigg]  \bigg\} \nonumber \\
  && \qquad =  \;-\sum_{v \in I^\sigma} \PP_0(v) \ln[\G_0(v)] +  \lim_{\tau \rightarrow \infty} \bigg\{ \sum_{v \in I^\sigma} \PP_0(v) 
  		\ln \bigg[\sum_{v' : f^{\tau}(v') = \phi(v)} \G_0(v') \bigg]    \bigg\}  			\nonumber \\
   && \qquad =  \; \sum_{v \in I^\sigma} \PP_0(v)  \ln \bigg[ \frac{ \lim_{\tau \rightarrow \infty} \sum_{v' \in I^\sigma : N(v') \le \tau }
   		 \G_0(v')  }  {\G_0(v)} \bigg]      	\nonumber \\		
   && \qquad =  \; \sum_{v \in I^\sigma} \PP_0(v)  \ln \bigg[ \frac{\G_0(I^\sigma) }  {\G_0(v)} \bigg]  		 
\ea
where the penultimate step uses the fact that $\phi(v)$ is a fixed point for all $v \in I^\sigma$, and the last step uses the fact
that all $v \in I^\sigma$ eventually halt.

So defining expected work using this second approach,
the optimal $\PP_0$ is a delta function about the $v \in I^\sigma$ that minimizes $\G_0(v) \propto 2^{-\ell(v)}$. But that
is just the $v_0$ of minimal length in the set of all $v_0$ that result in output $\sigma$. The associated minimal expected
work is 
\ba
kT \ln(2)\bigg(K_U(\sigma) + \log[\G_0(I^\sigma)] + \log \lambda \bigg)
\ea
as claimed in~\erf{eq:kolmogorov}.

\end{document}